\documentclass[a4paper,english,final]{article}
\pdfoutput=1
\usepackage{amsmath,amscd}
\usepackage{amssymb}
\usepackage{amsthm}
\usepackage{babel}
\usepackage{color}
\usepackage{procalg}
\usepackage{url}
\usepackage{graphicx}
\usepackage{subfigure}
\usepackage[numbers]{natbib}

\theoremstyle{plain}
\newtheorem{theorem}{Theorem}[section]

\theoremstyle{definition}
\newtheorem{definition}[theorem]{Definition}

\newcommand{\com}{\ensuremath{?\mkern-6mu!}}

\newcommand{\chan}{\ensuremath{\mathcal{H}}}
\newcommand{\data}{\ensuremath{\mathcal{D}}}
\newcommand{\hide}{\ensuremath{\mathcal{I}}}
\newcommand{\block}{\ensuremath{\mathcal{B}}}

\newcommand{\br}{\ensuremath{\mathbf{Br}}}

\title{Design of asynchronous supervisors}
\author{Harsh Beohar \thanks{\protect\url{h.beohar@tue.nl}}
\and Pieter Cuijpers \thanks{\protect\url{P.J.L.Cuijpers@tue.nl}}
 \and Jos Baeten \thanks{\protect\url{josb@win.tue.nl}}}
\begin{document}

\maketitle

\begin{abstract}
One of the main drawbacks while implementing the interaction between a plant and a supervisor, synthesised by the supervisory control theory of \citeauthor{RW:1987}, is the inexact synchronisation. \citeauthor{balemiphdt} was the first to consider this problem, and the solutions given in his PhD thesis were in the domain of automata theory. Our goal is to address the issue of inexact synchronisation in a process algebra setting, because we get concepts like modularity and abstraction for free, which are useful to further analyze the synthesised system. In this paper, we propose four methods to check a closed loop system in an asynchronous setting such that it is branching bisimilar to the modified (asynchronous) closed loop system. We modify a given closed loop system by introducing buffers either in the plant models, the supervisor models, or the output channels of both supervisor and plant models, or in the input channels of both supervisor and plant models. A notion of desynchronisable closed loop system is introduced, which is a class of synchronous closed loop systems such that they are branching bisimilar to their corresponding asynchronous versions. Finally we study different case studies in an asynchronous setting and then try to summarise the observations (or conditions) which will be helpful in order to formulate a theory of desynchronisable closed loop systems.
\end{abstract}

\section{Introduction}\label{intro}
Supervisory control theory (RW-theory) \citep{RW:1987,RW:1989} performs automatic synthesis of a supervisor which controls a plant such that a corresponding requirement (legal behaviour) is achieved. In control theory terminology,
\begin{itemize}
\item the model which is to be controlled is known as \textit{plant},
\item the model which specifies the requirement is known as \textit{specification},
\item the model which forces the plant to meet the specification by interacting with it is known as \textit{supervisor}.
\item the interaction between the plant and the supervisor is known as \textit{closed-loop behavior}.
\end{itemize}
The closed loop behaviour in RW-theory is realized by synchronous parallel composition. Informally, it allows a plant and a supervisor to synchronise on common events while other events can happen independently.

One of the main drawbacks while implementing the interaction between a plant and a supervisor, synthesised by the supervisory control theory of \citeauthor{RW:1987}, is the inexact synchronization \citep{fabian}. In practical industrial application the interaction between a plant and a supervisor is not synchronous but rather asynchronous. Due to the synchronous parallel composition, the interaction between the plant and the supervisor is strict. By strict we mean that, either plant or supervisor has to wait for the other party while synchronising. To overcome this problem it is important to study asynchronous communication between the plant and the supervisor where communications are delayed in buffers. The choice of buffers depends on the domain of the system to be modeled. For instance, to model delay insensitive circuits, a wire (see \citep{diccs}) could be chosen as a buffering mechanism, while to model data-flow networks (see \citep{HHJ90}) a queue could be used as a buffering mechanism.

\citeauthor{balemiphdt} was the first to consider the inexact synchronisation problem, and the solutions given in his PhD thesis \citep{balemiphdt} were in the domain of automata theory. In \citep{balemiphdt} an \textit{input-output} interpretation was given between a plant and a supervisor and a special delay operator was introduced to model the delay in communication. Furthermore, to achieve modularity and abstraction in supervisory control theory, the original theory was extended with the concepts of decentralised control and partial observation, respectively. These concepts were also developed in \cite{balemiphdt}.

The disadvantage of a theory to be based on automata theory is that it requires development of some special concepts like decentralised control for modularity in case of RW-theory. If the theory is based on process algebra, these additional concepts can be attained for free. Congruence is one of the key features in process algebra which helps in achieving this modularity in system design. Modularity is a way of designing a complex system by dividing it into different smaller components or subsystems.

Process algebra is one of the ways in which one can formally specify a system behaviour. It contains different constructs such as sequential composition, and parallel composition which are used as a basic building block to specify any desired behaviour. Apart from this, process algebra also provides modularity in system design, and abstraction from behaviour in system analysis. In this paper we show that it is possible to redesign a synchronous closed loop system in in an asynchronous setting by performing three case studies. Finally, some conditions are given which will be helpful in formulating the theory of desynchronisable closed loop systems. A synchronous closed loop system is called desynchronisable iff it is branching bisimilar to a corresponding asynchronous closed loop system.

This report is organized as follows. Section~\ref{prelim} introduces the overall background required for this report and consists of three sub-sections with the following description. In subsection~\ref{spec-lang} we introduce the formal language in which different models (like plant, supervisor or requirement) are specified. In subsection~\ref{buffermodels} we define different buffer models in the specification language. In subsection~\ref{rwtheory} we give a brief introduction on RW-theory and discuss the relation and some results from previous literature with respect to our specification language. In section~\ref{sec-casestudies} the work-flow for the different case-studies is given. In particular, it explains how different tools can be used to refine synchronous communication into asynchronous communication. Then the subsections~\ref{subsec-2machbuf},\ref{subsec-pl}, and~\ref{subsec-pc} are devoted to explain three different case studies. Finally, in section~\ref{sec-disc} we discuss the key conditions which were satisfied by the synchronous closed loop systems in the case studies such that they were branching bisimilar to the corresponding asynchronous version.

\section{Preliminaries}\label{prelim}
\subsection{Specification language}\label{spec-lang}
We consider the TCP process algebra \citep{acpbook} as the suitable formalism which will be used throughout this paper. This choice is motivated by the following two main reasons:
\begin{itemize}
\item One of our goals is to develop this theory and implement it in the current $\chi$ \cite{chi} tool set as an extended functionality. We know that TCP as a language is a subset of $\chi$ and is simpler to work upon, as the latter has a constructs to model hybrid systems while the former is used to model discrete event systems only.
\item Previous studies \cite{HHJ90,Fischer96,Pursuit} of an asynchronous process composed using buffers used failure equivalence. By studying asynchronous system using TCP we want to find out whether it is possible to state results in a equivalence finer than trace or failure equivalence (see \citep{Glabeek90}, for the lattice of process equivalences).
\end{itemize}

In this paper we use $\data$ and $\chan$ to denote finite sets of data elements and channel names, respectively. Then for each channel $h\in \chan$ and each $d\in\data$, assume the presence of the following atomic actions:
\begin{itemize}
\item $h!d$ : send a data element $d$ at channel $h$,
\item $h?d$ : receive $d$ at channel $h$, and
\item $h\com d$ : communicate $d$ at channel $h$.
\end{itemize}

The following notation and definition will be used throughout the paper. The complete set of actions is denoted by $\Act$ where $\Act=\{h!d,h?d,h\com d\mid \forall d\in\data \wedge \forall h\in\chan\}$. Then the communication function $\gamma:\Act\times\Act\rightarrow\Act$ is defined for all $h\in\chan$ as $\gamma(h!d,h?d)=\gamma(h?d,h!d)=h\com d$ and undefined otherwise. Define the blocking set as $\block=\{h?d,h!d\mid d\in\data \wedge h\in\chan\}$ and the hiding set as $\hide=\{h\com d\mid d\in\data\wedge h\in\chan\}$.
The set of all process terms (denoted by $\Prc$) is then defined by the following grammar:
\begin{center}
$\begin{array}{lcll}
    \Prc& \triangleq & \dl & \mbox{deadlock process}\\
    &\mid & \emp & \mbox{empty process or \textit{skip}}\\
    &\mid & h\textbf{\large{!}}d\seqc \Prc & \mbox{action prefix, where }\textbf{\large{!}}\in\{?,!,\com\}\\
    &\mid & \Prc\altc\Prc & \mbox{alternative composition}\\
    &\mid & \Prc\merge\Prc & \mbox{parallel composition}\\
    &\mid & \encap{B}{\Prc} & \mbox{action encapsulation, where }B\subseteq\block\\
    &\mid & \abstr{I}{\Prc} & \mbox{abstraction (hiding of actions), where }I\subseteq\hide\\
    &\mid & \rname{f}{\Prc} & \mbox{renaming of process, where} f:\Act\rightarrow\Act.\\
    &\mid & \mathcal{R} & \mbox{recursive definition}
\end{array}$
\end{center}
The notation $\mathcal{R}$ denotes a recursion definition by a set of pairs $\{X_0=t_0,\dots, X_m=t_m\}$ where $X_i$ denotes a recursion variable and $t_i$ the process term defining it. The formal semantics for these operators can be found in \citep{acpbook}.

Note that in the above definition of process terms it is possible that a process may have the same channel for sending and receiving a data element. But such processes causes a problem while constructing an asynchronous closed loop system from its synchronous counter part. The problem is following: "Suppose a process $X$ has $h?a$ and $h!b$ in its alphabet. Then the information whether $h$ is an input or an output channel is unknown. In this paper we construct an asynchronous process by introducing input queues in input channels and output queues in output channels. The difference between input and output queues will be cleared later in Section~\ref{sec-casestudies}. Thus, the information whether an input or an output queue should be attached with channel $h$ becomes unclear".

In order to simplify things, we assume that every process has different channels for sending and receiving. Let $\alpha(Q)$ denote the alphabet (see \citep{acpbook}) of a process $Q$. Formally, a process $Q\in\Prc$ is called a simple process iff $\forall h,d.[ h!d\in\alpha(Q)\Rightarrow \not\exists d'.[h?d'\in\alpha(Q)]$ and vice versa. We will work with plants and supervisors which are specified as simple processes. 

We now introduce the transition system for a process and for a synchronous closed loop system which will be helpful in defining conditions, presented in Section~\ref{sec-disc}. A transition system generated by a process $X\in\Prc$ is denoted by quintuple $T_X=(Q_X,\rightarrow_X,q^i_X,A_X)$, where $Q_X$ denotes the set of states, $\rightarrow_X\subseteq Q_X\times A_X\times Q_X$ is the transition relation, $q^i_X$ is the initial state of the process $X$, and $A_X\subseteq\Act$ is the alphabet of $X$. In this paper we make distinction between a transition system for a process and a transition system for synchronous ( or asynchronous ) closed loop system. This distinction is based on the alphabet of a process, i.e. for a process $X$ which can be used to model a plant or a supervisor the alphabet of $X$ should be a set $B\subseteq\block$ while for a closed loop system $Y$ it should be a set $I\subseteq\hide$. We use notation $T_X$ for a transition system of a process X, $T_{SC}$ for a synchronous closed loop system. In the next subsection we define the different buffer processes which will be used later to study asynchronous communication with respect to these different kinds of buffer.

\subsection{Buffer models}\label{buffermodels}
In the previous subsection we defined the syntax of the formal language which will be used for specifying plant, supervisor, requirement and buffers. A buffer is a process which receives data from another process and stores that data until another process reads it. For example, a buffer can be a queue (FIFO), or a stack (LIFO), or a wire, or a bag. Before defining the different buffer processes in the above language, we need to define an auxiliary set for channels. This is necessary for the conversion of a given synchronous closed system into an asynchronous one. So assume that the set $\chan$ is closed under $\hat{\_}\:$, i.e. if $h\in\chan$ then also $\hat{h}\in\chan$. Then define renaming functions $\hat{f}:\chan \rightarrow \chan$ and $f:\chan\rightarrow\chan$ as follows:
\begin{itemize}
\item for any $k\in\chan, \hat{f}(k)=\hat{k}$,
\item for any $\hat{k}\in\chan, f(\hat{k})=k$,
\item for $k\in\chan$ not in the image of $\hat{f}$, $f(k)=k$.
\end{itemize}
The subscript notation $f_i$ ($f_o$) is used to indicate the renaming of input (output) channels only. Now we give the formal definition for different types of buffers.
\begin{definition}{(\textbf{Queue}).}\label{queuedef}
Let $\varepsilon$ denote the empty list. Let $\xi$ denote a list of data elements. Let $e.\xi$, and $\xi.d$ denote a list with first element $e$ and last element $d$, respectively. Then, a queue with input channel $h\in \chan$ and output channel $\hat{h}=\hat{f}(h)$ is specified as follows:
\begin{eqnarray*}
\mathcal{Q}_{h}(\varepsilon)&=&\sum_{d\in \data}h?d\seqc \mathcal{Q}_{h}(d.\varepsilon)\\
\mathcal{Q}_{h}(\xi.d) &=& \hat{h}!d\seqc \mathcal{Q}_{h}(\xi)+\sum_{e\in \data}h?e\seqc \mathcal{Q}_{h}(e.\xi.d)\\
&&\mbox{ (for every $\xi\in\data^{*},d\in\data$.)}
\end{eqnarray*}
Now define a queue for every set of channels $H \subseteq \chan$:\newline
\[\mathit{Queue}_{H}= \merge_{h \in H}\mathcal{Q}_{h}(\varepsilon) \hspace{7.7cm}\qed \]
\end{definition}
\begin{definition}{(\textbf{Stack}).}\label{buffdef}
A stack with input channel $h\in \chan$ and output channel $\hat{h}=\hat{f}(h)$ is specified as follows (with parameters as in definition~\ref{queuedef}):
\begin{eqnarray*}
\mathcal{S}_{h}(\varepsilon)&=&\sum_{d\in \data}h?d\seqc \mathcal{S}_{h}(d.\varepsilon)\\
\mathcal{S}_{h}(d.\xi) &=& \hat{h}!d\seqc \mathcal{S}_{h}(\xi)+\sum_{e\in \data}h?e\seqc \mathcal{S}_{h}(e.d.\xi)\\
&&\mbox{ (for every $\xi\in\data^{*},d\in\data$.)}
\end{eqnarray*}
Now define a stack for every set of channels $H \subseteq \chan$:\newline
\[\mathit{Stack}_{H}= \merge_{h \in H}\mathcal{S}_{h}(\varepsilon) \hspace{7.7cm}\qed \]
\end{definition}
\begin{definition}{(\textbf{Wire}).}\label{buffdef}
A wire with input channel $h\in \chan$ and output channel $\hat{h}=\hat{f}(h)$ is specified as follows (with parameters as in definition~\ref{queuedef}):
\begin{eqnarray*}
\mathcal{W}_{h}(\varepsilon)&=&\sum_{d\in \data}h?d\seqc \mathcal{W}_{h}(d.\varepsilon)\\
\mathcal{W}_{h}(d)&=&\hat{h}!d\seqc\mathcal{W}_{h}(d)+\sum_{e\in\data}h?e\seqc\mathcal{W}_{h}(e).
\end{eqnarray*}
Now define a wire for every set of channels $H \subseteq \chan$:\newline
\[\mathit{Wire}_{H}= \merge_{h \in H}\mathcal{W}_{h}(\varepsilon) \hspace{7.7cm}\qed \]
\end{definition}
\begin{definition}{(\textbf{Bag}).}\label{buffdef}
Let $\emptyset$ denote the empty multiset. Let $\xi$ denote a multiset of data elements . Let $\xi\Cup\{e\}$ denote the multiset $\xi$ with the multiplicity of $e$ increased by $1$ and $\xi\Cap\{e\}$ denote the multiset $\xi$ with the multiplicity of $e$ decreased by $1$. Then, a bag with input channel $h\in \chan$ and output channel $\hat{h}=\hat{f}(h)$ is specified as follows:
\begin{eqnarray*}
\mathcal{B}_{h}(\emptyset)&=&\sum_{d\in \data}h?d\seqc \mathcal{B}_{h}(\emptyset\Cup\{d\})\\
\mathcal{B}_{h}(\xi) &=& \sum_{e\in \xi}\hat{h}!e\seqc \mathcal{B}_{h}(\xi\Cap \{e\})+\sum_{f\in \data}h?f\seqc \mathcal{B}_{h}(\xi\Cup\{f\})\\
&&\mbox{ (for every $\xi\in\data^{*},d\in\data$.)}
\end{eqnarray*}
Now define a bag for every set of channels $H\subseteq \chan$:\newline
\[\mathit{Bag}_{H}= \merge_{h \in H}\mathcal{B}_{h}(\emptyset) \hspace{7.7cm}\qed \]
\end{definition}

\subsection{Supervisory control theory}\label{rwtheory}
In this subsection we give a brief introduction to the RW-theory in our setup. The basic building block in RW-theory is a deterministic automaton. Plants and supervisors are allowed to perform actions or events which are divided into two disjoint subsets: controllable events and uncontrollable events, i.e. $\data=\data_c\uplus\data_{uc}$. The idea behind this partition is that the supervisor can enable or disable controllable events so that the closed loop behavior is the same as the specification under language equivalence. Furthermore, it can observe but cannot influence uncontrollable events.

\begin{figure}
  \centering
  \includegraphics[width=6cm,bb=0 0 410 306]{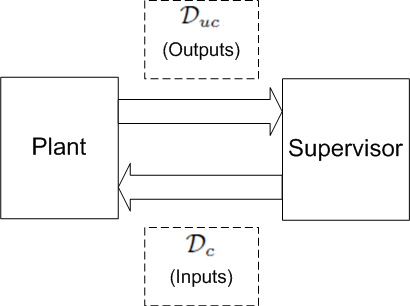}\\
  \caption{Context diagram for plant and supervisor.}\label{context}
\end{figure}

The two basic differences from the original theory and the current setup are following. Firstly, we use processes as the building blocks instead of automata. As a consequence we work with finer equivalence than language equivalence. Secondly, we follow the input-output interpretation \cite{balemiphdt} between a plant and a supervisor (see Figure~\ref{context}). In this interpretation the uncontrollable events are outputs from a plant to a supervisor and the controllable events are outputs from a supervisor to a plant.

Next we introduce the term deterministic process which will be helpful in defining a plant, supervisor and requirement models in our setup.

\begin{definition}
A process $Y\in\Prc$ is called a deterministic process \citep{acpbook} if and only if for all states $Y$ of the transition system (generated by the operational rules) it holds that $Y\step{a}U \wedge Y\step{a}Z \Rightarrow U\equiv Z$, where $U,Z\in\Prc$, and $U\equiv Z$ means $U$ and $Z$ are syntactically equivalent.
\end{definition}

The three basic entities in the RW-theory are: a plant, a supervisor, and a requirement. A plant is a simple and deterministic process $P\in\Prc$ that does not contain communication actions. The requirement of determinism is necessary because the RW-theory (and its synthesis tool TCT \citep{W:2008}) is based only on deterministic finite automata. The condition that a plant process does not contain communication actions can be stated formally as $\encap{\hide}{P}\bisim P$. The requirement of the above condition is due to the construction of asynchronous processes which is explained in detail in following lines. Buffers (like queues, stack, etc) are introduced in both input and output channel. So if communication actions (like $h\com d\in I$) are allowed in the specification of a plant process then the information whether the channel $h$ would be an input or an output channel of the plant process is unknown. Similarly, a supervisor is a simple and deterministic process $S\in\Prc$ such that $\encap{\hide}{S}\bisim S$.

A requirement is a process which specifies the legal interaction that should occur while the plant and supervisor are interacting such that a required task (for which supervisor is synthesised) is completed. Thus, a requirement is a deterministic process $E\in\Prc$ such that $\encap{\block}{E}\bisim E$. This condition suggests that a requirement process should contain only communication actions in its alphabet.

\begin{figure}\centering
  \subfigure[Synchronous closed loop system.]{\label{synccl}
\includegraphics[width=0.4\textwidth,bb=14 14 339 167]{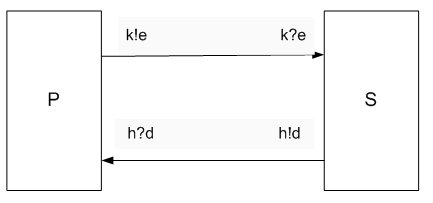}}
\subfigure[Asynchronous closed loop system.]{\label{asynccl}
\includegraphics[width=0.8\textwidth,bb=14 14 941 346]{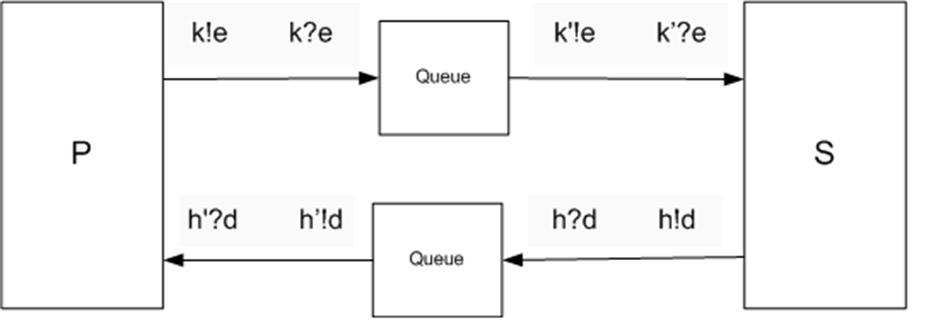}
  }
\caption{Illustration of the research question.}\label{cm}
\end{figure}

Now we can state the control problem as follows: find a supervisor $S$ for a given plant $P$ and a given requirement $E$ such that, \[\encap{\block}{P\merge S}\bisim E.\] In this paper we do not consider how the supervisor is computed and rather use the solution \citep{W:2008} which provides a closed loop system $\encap{B}{P\merge S}$ strongly bisimilar to a requirement $E$. Then the aim of this paper is to check whether it is possible to construct an asynchronous closed loop system (Figure~\ref{asynccl}) such that it is branching bisimilar with its corresponding synchronous closed loop system (Figure~\ref{synccl}).

\section{Approach}\label{sec-casestudies}
As already discussed in the introduction section, it is important to study an asynchronous interaction between a plant and a supervisor to model an industrial application. So in this section we first give four ways to construct an asynchronous closed loop system from a given synchronous closed loop system. Then, a work-flow is presented which explains how different tools can be used to refine synchronous communication into asynchronous communication in a straightforward and correct way. Later we redesign three case studies in an asynchronous setting based upon the above work-flow and methods which were already modelled by the System Engineering group of Eindhoven university of technology (TU/e) in the synchronous setting \cite{4k460}.

\begin{figure}\centering
  \subfigure[Construction method, M1.]{\label{cm-m1}
\includegraphics[width=0.7\textwidth,bb=14 14 432 197]{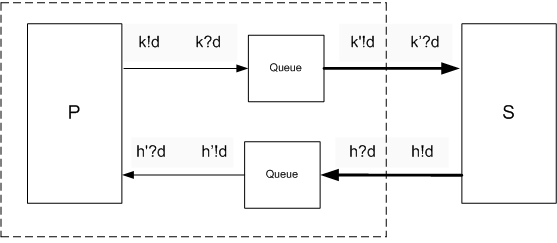}}
\hspace{1cm}  \subfigure[Construction method, M2.]{\label{cm-m2}
\includegraphics[width=0.7\textwidth,bb=14 14 435 199]{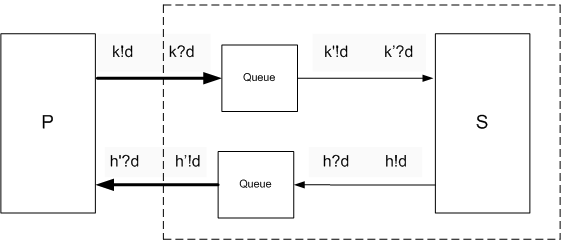}
  }
\hspace{1cm} \subfigure[Construction methods, M3.]{\label{cm-m3}
\includegraphics[width=0.7\textwidth,bb=14 14 447 201]{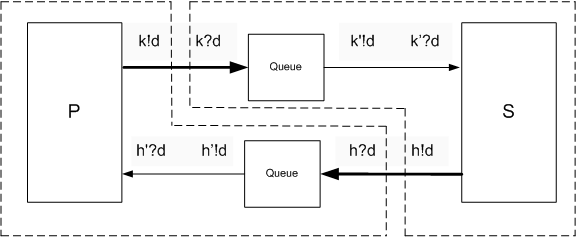}
  }
\hspace{1cm} \subfigure[Construction methods, M4.]{\label{cm-m4}
\includegraphics[width=0.7\textwidth,bb=14 14 456 198]{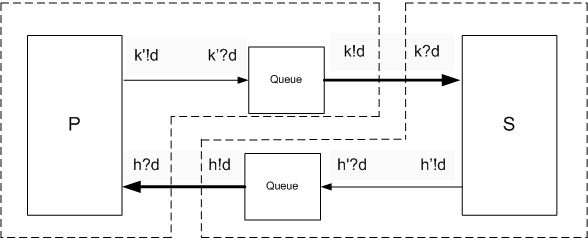}
  }
\caption{Different ways to construct an asynchronous closed loop system.}\label{cm}
\end{figure}

A synchronous closed loop system (Figure~\ref{synccl}) can be converted into an asynchronous one by introducing queues (see Fiugre~\ref{asynccl}) in following ways:
\begin{itemize}
\item[M1.] introducing queues between the plant and supervisor process models such that \textit{the interaction between the plant and queues are hidden} (see Figure~\ref{cm-m1}). The thick lines are used to indicate the visible interaction and thin lines are used to indicate the invisible interaction in Figure~\ref{cm}.
\item[M2.] introducing queues between the plant and the supervisor process models such that \textit{the interaction between the supervisor and queues are hidden} (see Figure~\ref{cm-m2}).
\item[M3.] introducing the queues between a plant and a supervisor such that \textit{the interaction between the output channels of both plant and supervisor with their corresponding queues are hidden} (see Figure~\ref{cm-m3}).
\item[M4.] introducing the queues between a plant and a supervisor such that \textit{the interaction between the input channels of both plant and supervisor with their corresponding queues are hidden} (see Figure~\ref{cm-m4}).
\end{itemize}
Note that the above indices {M1, M2, M3, and M4} are important as they will be used while presenting the results obtained from all the three case studies.

\begin{figure}
\begin{center}
  \includegraphics[width=1\textwidth,bb=14 14 758 533]{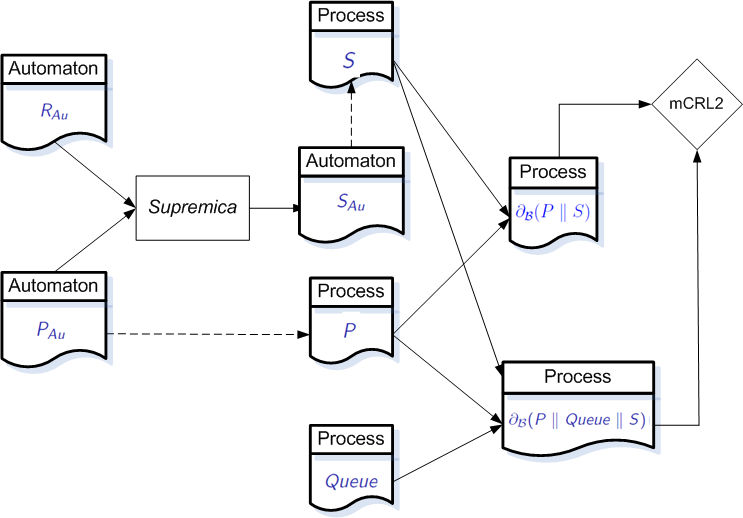}
\caption{Work-flow for refinement of synchronous communication into asynchronous communication}\label{workflow}
\end{center}
\end{figure}

To study an asynchronous interaction between a plant and a supervisor we present the work-flow shown in Figure~\ref{workflow}, which checks whether the synchronous and the asynchronous implementation of a plant and a supervisor are equivalent. We check for both branching bisimulation equivalence and weak trace equivalence between the two closed loop systems. Our approach assumes that a plant model and a requirement model are given as automata. The tool \textit{Supremica} \citep{supremica} is used to synthesise the supervisor for a given plant and a given requirement model. The synthesised automaton is then converted into a process algebraic model in TCP language. Similarly the plant automaton model is also converted into a process algebraic model. This conversion of automaton models into process models is done manually, indicated by the dashed lines (Figure~\ref{workflow}). Finally, the tool-set \textit{mcrl2} \citep{mcrl2} is used to check for the branching bisimulation relation between a synchronous closed loop and an asynchronous closed loop system designed by each construction method. The following case studies are modified in this report under asynchronous setting:
\begin{enumerate}
\item Two machines and a buffer example.
\item Pusher-lift system.
\item Pneumatic cylinder.
\end{enumerate}
For each of the case studies we follow the work-flow as presented. In the next subsections we first introduce the three case studies, and in the last subsection~\ref{subsec-results} we present the overall results obtained in a table. The mCRL2 specification for all the three case studies can be found in Appendix~\ref{appendix-2machbuf},\ref{appendix-pl},\ref{appendix-pc}.

\begin{figure}\centering
  \subfigure[Plant Models.]{\label{2mach-plant}
\includegraphics[width=0.4\textwidth,bb=14 14 348 104]{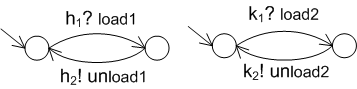}}
\hspace{1cm}  \subfigure[Requirement model.]{\label{2mach-req}
\includegraphics[width=0.2\textwidth,bb=14 14 176 107]{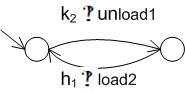}
  }
\vspace{1cm} \subfigure[Supervisor model.]{\label{2mach-sup}
\includegraphics[width=0.5\textwidth,bb=14 14 501 477]{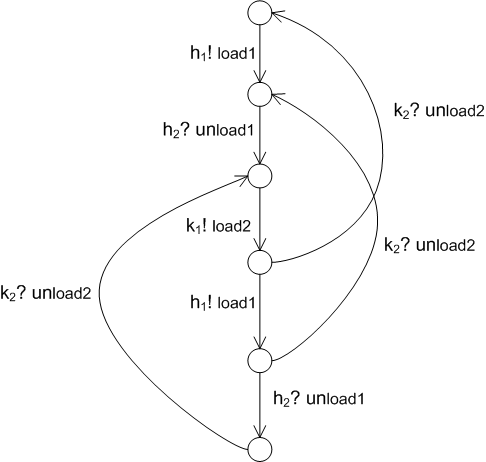}
  }
\caption{Two machines and a buffer example.}\label{2machbuf}
\end{figure}
\subsection{Two machines and a buffer example.}\label{subsec-2machbuf}
This case study is adapted from the examples given in the Supremica tool set \cite{supremica}. The case study consists of two machines which are connected through a buffer. The control task is to synthesise a supervisor which controls the two machines such that the requirement is met, see Figure~\ref{2mach-req}. The plant models are shown in Figure~\ref{2mach-plant}, the requirement model in Figure~\ref{2mach-req} and the synthesised supervisor in Figure~\ref{2mach-sup}. Note that the synchronous closed loop system for this case study is isomorphic to the supervisor transition system, except for the naming of the action labels. The action labels in a closed loop system will have $\com$ symbol, while in a supervisor process they will be annotated with either $?$ or $!$.

The asynchronous system contained finitely many states and the results pertaining to this case study are shown in Figure~\ref{results}.

%
\begin{figure}\centering
\subfigure[Pusher-lift system \cite{4k460}.]{\label{pusherlift}\includegraphics[width=8cm,bb=14 14 868 454]{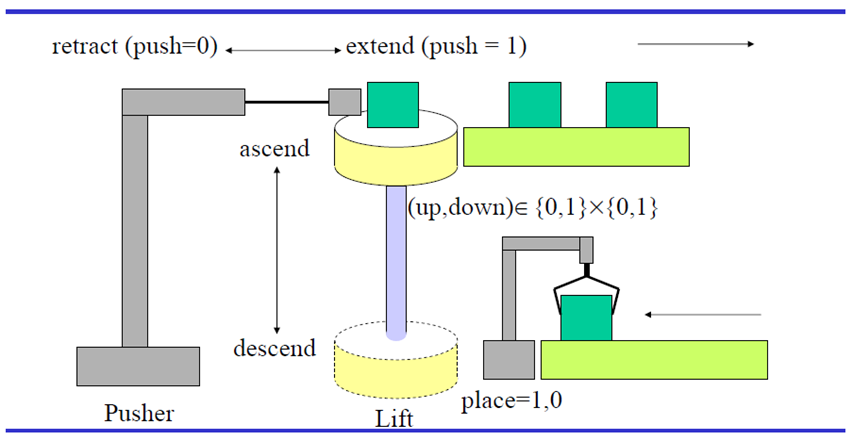}}
\subfigure[Pusher and Product model]{\label{pushlift-pushprod}\includegraphics[width=.5\textwidth, bb=14 14 498 235]{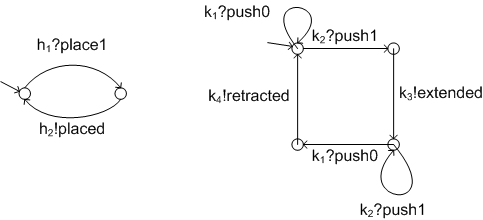}}
\subfigure[Lift model]{\label{pushlift-lift}\includegraphics[width=0.6\textwidth, bb=14 14 602 869]{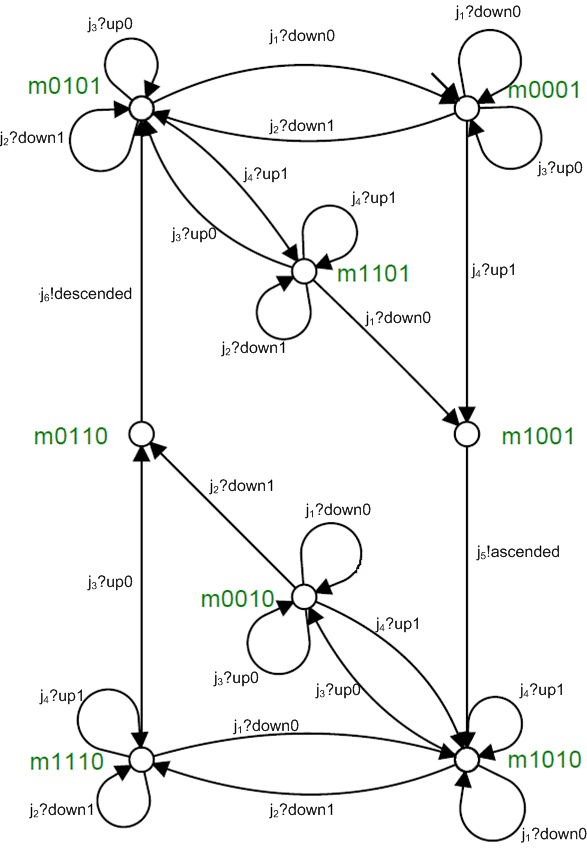}}
\caption{Plant models for Pusher-lift system.}\label{plantmodels-pl}
\end{figure}

\subsection{Pusher-lift system \protect{\cite{4k460}}}\label{subsec-pl}
\begin{figure}\centering
\includegraphics[width=.7\textwidth, bb=14 14 625 481]{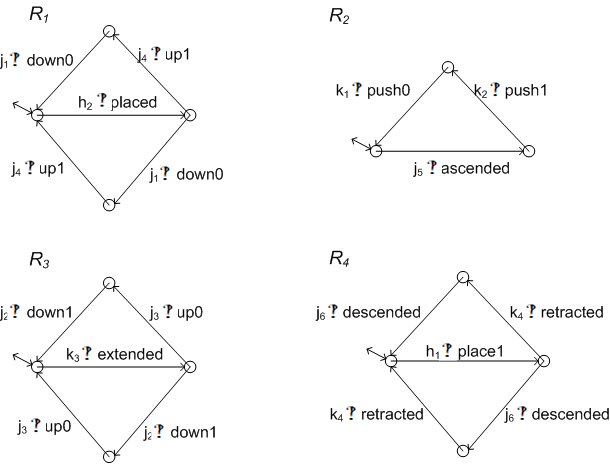}
\caption{Requirement models for Pusher-lift system.} \label{pushlift-req}
\end{figure}

\begin{figure}\centering
\includegraphics[width=.6\textwidth, bb=14 14 620 613]{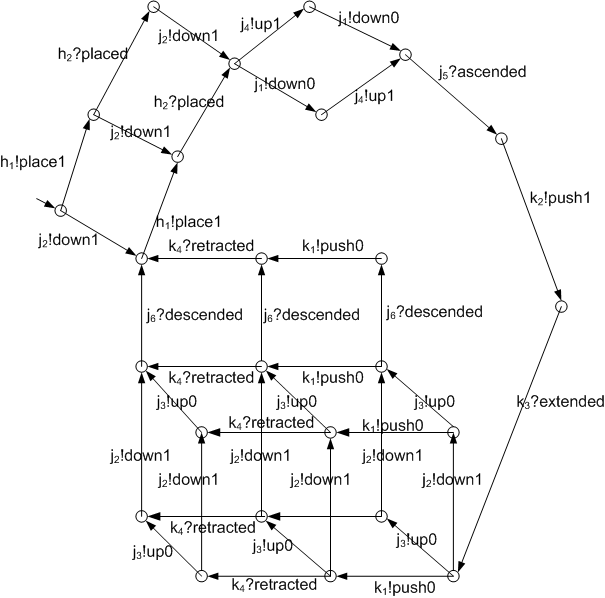}
\caption{Supervisor model for Pusher-lift system.}\label{sup-pl}
\end{figure}

The pusher lift system is a case study taken from a set of lecture notes on supervisory control course \cite{4k460}. The overall system consists of a lift that can go up and down, a pusher that can retract and extend, and a product holder (see Figure~\ref{pusherlift}). The plant model of the lift is shown in Figure~\ref{pushlift-lift}, pusher and product holder models in Figure~\ref{pushlift-pushprod}, and the different requirements are shown in Figure~\ref{pushlift-req}. The synthesised supervisor model using the Supremica tool is shown in Figure~\ref{sup-pl}. Note that the synchronous closed loop system for this case study is also isomorphic to the supervisor transition system, except for the naming of the action labels.


\begin{figure}\centering
  \includegraphics[width=0.8\textwidth,bb=14 14 728 308]{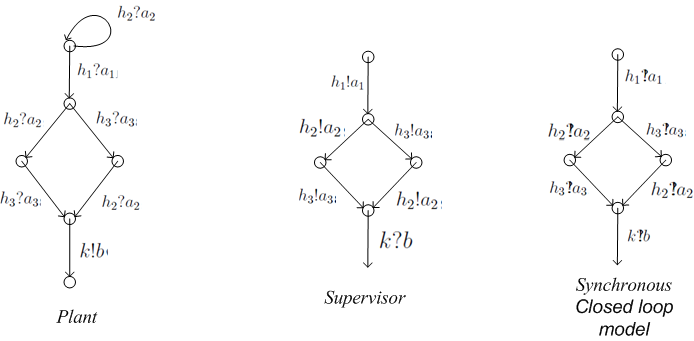}\\
  \caption{Deadlock caused by the self loop.}\label{restr1}
\end{figure}

The asynchronous closed loop system composed by all the four construction methods contained a deadlock for this case. Further analyzing the asynchronous closed loop system it was identified that the cause of the deadlock was a self loop in the plant model. This situation is explained by the following example in Figure~\ref{restr1} where a plant, a supervisor and a synchronous closed loop is given.

When the asynchronous closed loop system is designed with the construction method M1, the following trace indicated the deadlock: $<h_1\com a_1\seqc h_3\com a_3\seqc h_2\com a_2\seqc\hat{h}_2\com a_2\seqc\hat{h}_1\com a_1\seqc\hat{h}_3\com a_3>$ as the transition $k\com b$ is not possible. Note that in the above trace the actions decorated with ``$\;\hat{ }\;$'' will be performed by plant model. Moreover, the removal of self loop ($h_2\com a_2$) does not affect the synchronous closed loop system and then the above trace will not be valid for the modified asynchronous closed loop.

Note that the results obtained for all the construction methods are shown in Figure~\ref{results} with respect to both, modified plants (i.e. plant models without self loops) and original plants.

\begin{figure}\centering
\includegraphics[width=5cm]{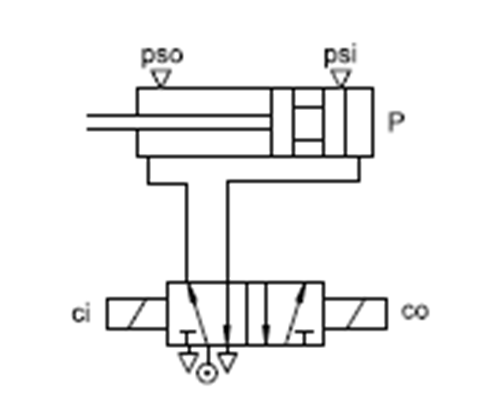}
\caption{Schematic diagram of pneumatic cylinder \protect{\cite{4k460}}.}\label{intro-pc}
\end{figure}

\subsection{Pneumatic cylinder \cite{4k460}}\label{subsec-pc}
\begin{figure}
\subfigure[Plant and requirement models for Pneumatic cylinder.]{\label{pcylinder}\includegraphics[width=6cm, bb=14 14 605 503]{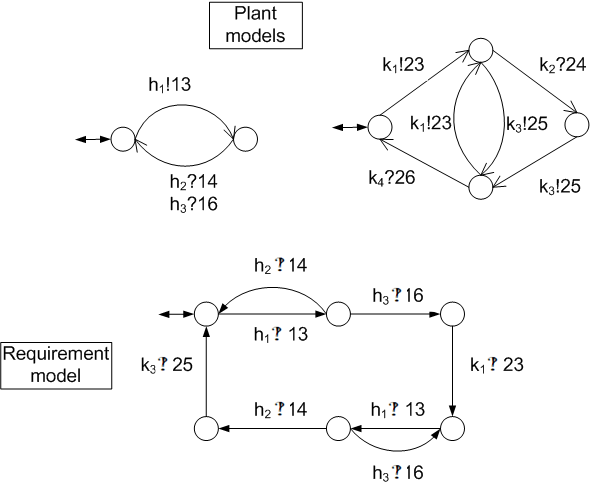}}\hspace{1cm}
\subfigure[Synchronous closed loop model (Supervisor model) for Pneumatic cylinder.]{\label{sup-pc}\includegraphics[width=8cm, bb=14 14 762 723]{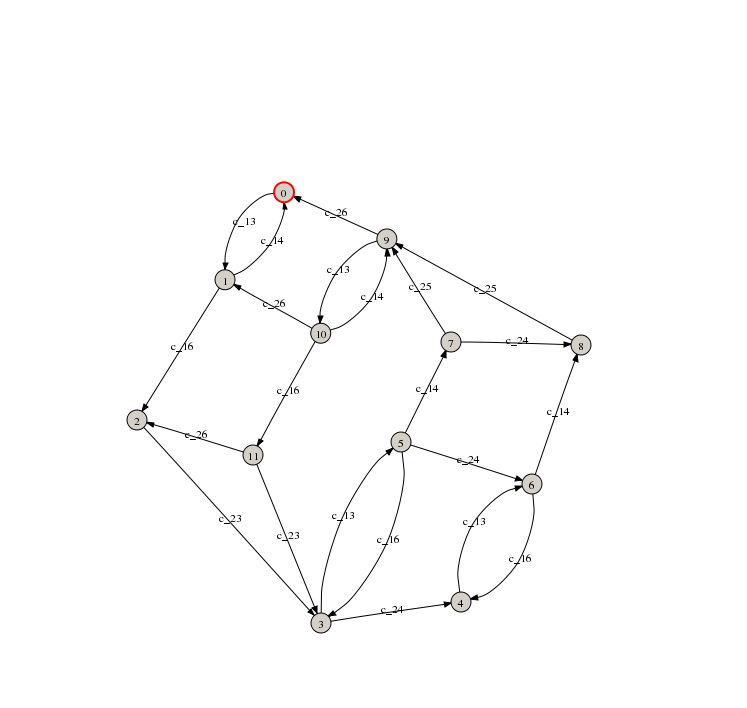}}
\caption{Pneumatic cylinder case study models.}\label{pneumatic}
\end{figure}

The task in this case study is to design a supervisor that makes a cylinder move out when a push button is activated.
Pushing the button will start the extending movement and releasing the button will start the retracting movement (see Figure~\ref{intro-pc}). The control signals $ci$ and $co$ are used to make the cylinder move in and out. The detection of the cylinder being at its innermost (outermost) position is realized through sensor $psi$ ($pso$). The plant and requirement models are shown in Figure~\ref{pcylinder}. The synthesised supervisor is shown in Figure~\ref{sup-pc}. Note that the transition system of synchronous closed loop system is again isomorphic to supervisor model, except for the action labelling.

In this case all the asynchronous closed loop systems designed by different methods were containing infinitely many states. To further analyse this system, all the asynchronous closed loop systems were redesigned with 1 place queues. The results pertaining to all the construction methods are shown in Figure~\ref{results}.

\subsection{Results}\label{subsec-results}
\begin{figure}\centering
\includegraphics[width=1.1\textwidth,bb=14 14 1048 517]{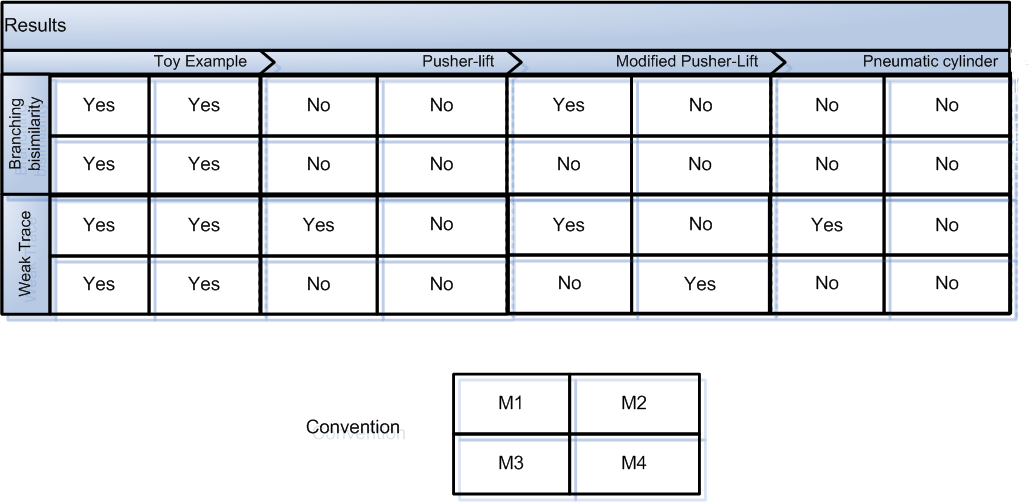}
\caption{Results obtained from all the case studies.}\label{results}
\end{figure}

In this subsection we present the results obtained for all the three case-studies in the Figure~\ref{results}. The following are the key observations found in the table shown in Figure~\ref{results}. We use the phrase `positive result' to mean that synchronous closed loop system is equivalent to asynchronous one.
\begin{itemize}
\item The construction method M1 yielded positive result for the all case studies with respect to weak trace equivalence.
\item In the modified pusher lift case study, only M1 and M3 yielded positive result under weak trace equivalent even though the asynchronous closed loop system constructed by M1 was branching bisimilar to synchronous one.
\item The toy example case study had positive results for all the construction methods.
\end{itemize}
The construction method M1 satisfied the most number of case studies under weak trace and branching bisimulation equivalence. This makes M1 a suitable candidate to further study and answer our research question.

\section{Discussion}\label{sec-disc}
In this section we sketch the conditions for a synchronous closed loop system to be desynchronisable. A synchronous closed loop system is called desynchronisable iff it is branching bisimilar to the corresponding asynchronous closed loop system. These conditions are important as they prevent the constructed asynchronous closed loop system from deadlocks and generation of infinitely many states. Before we give these conditions formally we need some auxiliary definitions. First, we ask the reader to recall the definitions of transition system of a process, and a synchronous closed loop system as defined in Section~\ref{spec-lang}.

Let $\eta:\mathcal{Q_{C}}\rightarrow 2^\hide$ be a function which returns a set of enabled actions at a state in a synchronous closed loop system $C$, i.e. $\eta(q_C)=\{a\mid q_C\step{a}\}$, where $q_C\in\mathcal{R},a\in\hide$. Let $\br:\mathcal{R}\rightarrow\mathbb{N}$ be a function defined as $\br(q_c)=|\eta(q_c)|$, which is used to access the degree of branching at a state.

Furthermore, we partition the set $\hide$ (defined in Section~\ref{spec-lang}) into two disjoint subsets $\hide_{P}^i,\hide_{P}^o$ with respect to the given plant process $P$ as:
\begin{itemize}
\item $\hide_{P}^i=\{h\com a\mid h\com a\in \hide\wedge h\in inch(P)\}$.
\item $\hide_{P}^o=\{k\com a\mid k\com a\in \hide\wedge k\in outch(P)\}$.
\end{itemize}

The following conditions are sufficient for desynchronisability:
\begin{enumerate}
\item Plant and supervisor composition must be \textit{well posed}. This term is borrowed from \cite{balemiphdt} where it was used for similar purpose, i.e. to ensure the asynchronous closed loop system is deadlock free. Consider the transition system $T_P=(Q_P,\rightarrow_P,q^i_P,\block)$ for a plant process $P$. Similarly consider $T_S$ as the transition system of a supervisor process $S$ with $T_S=(Q_S,\rightarrow_S,q^i_S,\block)$. Let $N:\block^{*}\rightarrow\block^{*}$ be a function defined as: $N(h?a.s)=h!a.N(s)$ and $N(h!a.s)=h?a.N(s)$ for some sequence $s\in\block^{*}$, and the dot symbol (.) indicates the concatenation of the sequence. Then, the plant and supervisor composition is called \textit{well posed} iff the following conditions are satisfied:
\begin{eqnarray*}
\forall s\in\block^{*},h!a\in\Act.[q^i_P\steps{s}_P q_P\step{h!a}_P&\Rightarrow& q^i_S\steps{N(s)}_S q_S\step{h? a}_S] \wedge \\
\forall s\in\block^{*},h!a\in\Act.[q^i_S\steps{s}_S q_S\step{h!a}_S&\Rightarrow& q^i_P\steps{N(s)}_P q_P\step{h?a}_P].
\end{eqnarray*}
For example, consider a plant process $P=h!a\seqc l?c\seqc P+k!b\seqc\dl$ and a supervisor process $S=h?a\seqc l!c\seqc S$. It is easy to verify that the synchronous closed loop system is deadlock free, as $\encap{\block}{P\merge S}=h\com a\seqc l\com c\seqc \encap{\block}{P\merge S}$. But when asynchronous closed loop system is designed using construction method M1 it will deadlock, because the plant can reach a deadlock state ($\dl$) by performing an action $k!b$.
\item No self loops in either plant model or supervisor model, i.e. both plant and supervisor should not contain a state such that a transition from that state lead into that same state. Let $T_P,T_S$ be the transition system for a plant and a supervisor, respectively. Then, $T_j$ must satisfy the condition $\forall a\in\block,\cnot\exists q_j\in Q_j.[q_j\step{a}q_j]$ for $j\in\{P,S\}$.

    The need of this condition was explained with an example in Pusher-lift case study (Section~\ref{subsec-pl}), which resulted in a deadlocked state in asynchronous closed loop system even though the synchronous closed loop system was deadlock free.
\begin{figure}\centering
\includegraphics[width=0.4\textwidth,bb=14 14 347 153]{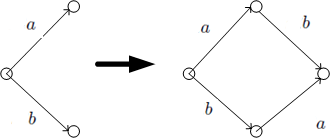}
\caption{Important property for desynchronisability.}\label{diamond-fig}
\end{figure}
\item All the states in a transition system for closed loop system must satisfy the diamond property (see Figure~\ref{diamond-fig}). Let $T_{C}$ be a transition system of a synchronous closed loop system $C$. Then $T_C$ is said to satisfy diamond property iff the following condition holds\begin{eqnarray*}
     \forall a,b\in\hide,q,q_1,q_2\in\mathcal{Q_C}.\Big[q\step{a}q_1\wedge q\step{b}q_2\Rightarrow \exists q_3.[q_1\step{b}q_3 \wedge q_2\step{a}q_3]\Big].\end{eqnarray*} This condition is required for establishing a branching bisimulation relation between a synchronous and an asynchronous closed loop system.
\item All the cycles from the initial state in a synchronous closed loop system must have at least one controllable, and one uncontrollable action in that cyclic trace. Again let $T_C$ be a transition system of a synchronous closed loop system. Then, \[\forall \pi\in\hide^{*}.\left[q^i_C\steps{\pi}q^i_C \Rightarrow (\pi\cap\hide_{P}^i\neq\emptyset)\wedge(\pi\cap\hide_{P}^o\neq\emptyset)\right].\]
Note, that we abuse the notation $x \cap A$ to denote the set of elements that occur both in the sequence $x$ and the set $A$. Intuitively, this condition ensures the progress of the components (i.e. plant and supervisor) in an asynchronous interaction. Consider the following example where $P=(k?b\seqc h?a + h?a\seqc k?b)\seqc P$ and $S=(k!b\seqc h!a\altc h!a\seqc k!b)\seqc S$. It is clear to see that $\encap{\block}{P\merge S}=(k\com b\seqc h\com a\altc h\com a\seqc k\com b)\seqc\encap{\block}{P\merge S}$, and there exists a cycle of actions $(h\com a\seqc k\com b)^{*}$ such that all actions are controllable actions. In this case the asynchronous closed loop system will have infinite states because supervisor $S$ can always send the controllable actions $k!b$, or $h!a$ to the unbounded queue, and the plant can wait forever to remove these actions from the queue (see Figure~\ref{cycleprob}). In figure~\ref{cycleprob} the two states with triangles indicate that the transitions from these two states are same as that of the black state. A similar example can also be given in which cycles contain only uncontrollable actions.
\begin{figure}\centering
\includegraphics[width=0.4\textwidth, bb=14 14 337 218]{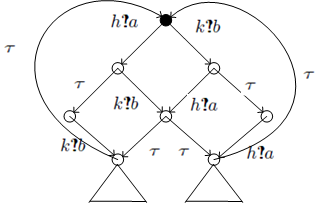}
\caption{Cycles containing only controllable actions causes infinite states.}\label{cycleprob}
\end{figure}
\end{enumerate}

Note that the formal proof of the above fact is under construction and will be published in future.

\paragraph{A Final Remark.} The transition systems generated by the four methods from a synchronous closed loop system are always isomorphic to each other apart from the difference in abstraction of actions. A hypothesis which one would expect to hold is that, all the four construction methods should always yield an equivalent asynchronous closed loop systems at least modulo weak trace equivalence. But the results shown in the Figure~\ref{results} implies that the above hypothesis is not true in general and the abstraction of actions does matter while reducing a transition system (the asynchronous one) even modulo weak trace equivalence. We conclude this report by framing the following open question namely, ``which of the construction methods classify a larger class of desynchronisable closed loop system with respect to an equivalence relation?''

\bibliographystyle{plainnat}
\bibliography{paper}
\appendix
\section{mCRL2 model for two machine and a buffer example}\label{appendix-2machbuf}
\begin{verbatim}
%%%%%%%%%%%%%Index for action names%%%%%%%%%%%%%%
%%l1 = load1
%%l2 = load2
%%ul1= unload1
%%ul2= unload2
%%%%%%%%%%%%%%%%%%%%%%%%%%%%%%%%%%%%%%%%%%%%%%%%%

act
s_l1,r_l1,s_l1',r_l1',c_l1,c_l1',s_l2,r_l2,s_l2',r_l2',c_l2,c_l2',
s_ul1,r_ul1,s_ul1',r_ul1',c_ul1,c_ul1',s_ul2,r_ul2,s_ul2',r_ul2',c_ul2,c_ul2';

proc

M1=r_l1'.s_ul1'.M1;
M2=r_l2'.s_ul2'.M2;

%M1=r_l1'.s_ul1.M1;
%M2=r_l2'.s_ul2.M2;

S0=s_l1.S4;
S4=r_ul1.S1;
S1=s_l2.S2;%
S2=r_ul2.S0 + s_l1.S5;
S5=r_ul2.S4 + r_ul1.S3;
S3=r_ul2.S1;

%S0=s_l1.S4;
%S4=r_ul1'.S1;
%S1=s_l2.S2;
%S2=r_ul2'.S0 + s_l1.S5;
%S5=r_ul2'.S4 + r_ul1'.S3;
%S3=r_ul2'.S1;

Ql1(x1: Int) = (x1==0)-> r_l1.Ql1(x1+1) <>
		 (s_l1'.Ql1(x1-1) + r_l1.Ql1(x1+1));

Ql2(x2: Int) = (x2==0)-> r_l2.Ql2(x2+1) <>
		 (s_l2'.Ql2(x2-1) + r_l2.Ql2(x2+1));

Qul1(ux1: Int) = (ux1==0)-> r_ul1'.Qul1(ux1+1) <>
		 (s_ul1.Qul1(ux1-1) + r_ul1'.Qul1(ux1+1));

Qul2(ux2: Int) = (ux2==0)-> r_ul2'.Qul2(ux2+1) <>
		 (s_ul2.Qul2(ux2-1) + r_ul2'.Qul2(ux2+1));

Plant=M1||M2;
Supervisor=S0;

init
hide({c_l1',c_l2',c_ul1',c_ul2'},
allow({c_l1,c_l2,c_ul1,c_ul2,c_l1',c_l2',c_ul1',c_ul2'},
    comm({s_l1|r_l1->c_l1,s_l2|r_l2->c_l2,s_ul1|r_ul1->c_ul1,
         s_ul2|r_ul2->c_ul2,s_l1'|r_l1'->c_l1',s_l2'|r_l2'->c_l2',
         s_ul1'|r_ul1'->c_ul1',s_ul2'|r_ul2'->c_ul2'},
        Plant||Ql1(0)||Ql2(0)||Qul1(0)||Qul2(0)||Supervisor    )))
  ;
\end{verbatim}
\section{mCRL2 model for Pusher-lift case study.}\label{appendix-pl}
\begin{verbatim}
%%%%%%Index for action names%%%%%%
%% asc=ascended
%% desc=descended
%% d0=down0
%% d1=down1
%% ext=extended
%% ret=retracted
%% pl1=place1
%% pld=placed
%% pu0=push0
%% pu1=push1
%% up0=up0
%% up1=up1
%%%%%%%%%%%%%%%%%%%%%%%%%%%%%%%%%%

act
  s_asc,r_asc,s_asc',r_asc',c_asc,c_asc',s_desc,r_desc,s_desc',r_desc',
  c_desc,c_desc',s_d0,r_d0,s_d0',r_d0',c_d0,c_d0',s_d1,r_d1,s_d1',r_d1',
  c_d1,c_d1',s_ext,r_ext,s_ext',r_ext',c_ext,c_ext',s_pl1,r_pl1,s_pl1',
  r_pl1',c_pl1,c_pl1',s_pld,r_pld,s_pld',r_pld',c_pld,c_pld',s_pu0,r_pu0,
  s_pu0',r_pu0',c_pu0,c_pu0',s_pu1,r_pu1,s_pu1',r_pu1',c_pu1,c_pu1',s_ret,
  r_ret,s_ret',r_ret',c_ret,c_ret',s_up0,r_up0,s_up0',r_up0',c_up0,c_up0',
  s_up1,r_up1,s_up1',r_up1',c_up1,c_up1';

proc
Pu1=r_pu0'.Pu1+r_pu1'.Pu2;
Pu2=s_ext'.Pu3;
Pu3=r_pu1'.Pu3 + r_pu0'.Pu4;
Pu4=s_ret'.Pu1;

%L0001=r_d0.L0001+r_up0.L0001+r_up1.L1001+r_d1.L0101;
L0001=r_up1'.L1001+r_d1'.L0101;

L1001=s_asc'.L1010;

%L1010=(r_up1+r_d0).L1010 + r_up0.L0010+r_d1.L1110;
L1010=r_up0'.L0010+r_d1'.L1110;

%L1110=(r_up1+r_d1).L1110 + r_d0.L1010 + r_up0.L0110;
L1110=r_d0'.L1010 + r_up0'.L0110;

%L0010=(r_up0+r_d0).L0010 + r_up1.L1010 + r_d1.L0110;
L0010=r_up1'.L1010 + r_d1'.L0110;

L0110=s_desc'.L0101;

%L0101=(r_up0+r_d1).L0101 + r_d0.L0001 +r_up1.L1101;
L0101=r_d0'.L0001 +r_up1'.L1101;

%L1101=(r_up1+r_d1).L1101 + r_up0.L0101+r_d0.L1001;
L1101=r_up0'.L0101+r_d0'.L1001;

Pr1=r_pl1'.Pr2;
Pr2=s_pld'.Pr1;

%Qpu0(pu0: Int) = (pu0==0)-> r_pu0.Qpu0(pu0+1) <>
%		 (s_pu0'.Qpu0(pu0-1) + r_pu0.Qpu0(pu0+1));

%Qpu1(pu1: Int) = (pu1==0)-> r_pu1.Qpu1(pu1+1) <>
%		 (s_pu1'.Qpu0(pu1-1) + r_pu1.Qpu1(pu1+1));

S0=s_d1.S11+s_pl1.S22;
S11=s_pl1.S9;
S22=s_d1.S9 + r_pld.S23;
S9=r_pld.S12;
S23=s_d1.S12;
S12=s_up1.S8 + s_d0.S24;
S8=s_d0.S21;
S24=s_up1.S21;
S21=r_asc.S2;
S2=s_pu1.S4;
S4=r_ext.S6;
S6=s_d1.S18+s_up0.S10+s_pu0.S3;
S18=s_up0.S20 + s_pu0.S15;
S10=s_d1.S20 + s_pu0.S7;
S3=s_d1.S15+s_up0.S7+r_ret.S1;
S20=r_desc.S16+s_pu0.S19;
S15=s_up0.S19+ r_ret.S14;
S7=s_d1.S19+r_ret.S5;
S1=s_d1.S14+s_up0.S5;
S16=s_pu0.S13;
S19=r_desc.S13+r_ret.S17;
S14=s_up0.S17;
S5=s_d1.S17;
S13=r_ret.S11;
S17=r_desc.S11;

Qpu0(x1: Int) = (x1==0)-> r_pu0.Qpu0(x1+1) <>
		 (s_pu0'.Qpu0(x1-1) + r_pu0.Qpu0(x1+1));

Qpu1(x2: Int) = (x2==0)-> r_pu1.Qpu1(x2+1) <>
		 (s_pu1'.Qpu1(x2-1) + r_pu1.Qpu1(x2+1));

Qext(x3: Int) = (x3==0)-> r_ext'.Qext(x3+1) <>
		 (s_ext.Qext(x3-1) + r_ext'.Qext(x3+1));

Qret(x4: Int) = (x4==0)-> r_ret'.Qret(x4+1) <>
		 (s_ret.Qret(x4-1) + r_ret'.Qret(x4+1));

Qd0(x5: Int) = (x5==0)-> r_d0.Qd0(x5+1) <>
		 (s_d0'.Qd0(x5-1) + r_d0.Qd0(x5+1));

Qd1(d1: Int) = (d1==0)-> r_d1.Qd1(d1+1) <>
		 (s_d1'.Qd1(d1-1) + r_d1.Qd1(d1+1));

Qup0(up0: Int) = (up0==0)-> r_up0.Qup0(up0+1) <>
		 (s_up0'.Qup0(up0-1) + r_up0.Qup0(up0+1));

Qup1(up1: Int) = (up1==0)-> r_up1.Qup1(up1+1) <>
		 (s_up1'.Qup1(up1-1) + r_up1.Qup1(up1+1));

Qasc(x6: Int) = (x6==0)-> r_asc'.Qasc(x6+1) <>
		 (s_asc.Qasc(x6-1) + r_asc'.Qasc(x6+1));

Qdesc(x7: Int) = (x7==0)-> r_desc'.Qdesc(x7+1) <>
		 (s_desc.Qdesc(x7-1) + r_desc'.Qdesc(x7+1));

Qpl1(pl1: Int) = (pl1==0)-> r_pl1.Qpl1(pl1+1) <>
		 (s_pl1'.Qpl1(pl1-1) + r_pl1.Qpl1(pl1+1));

Qpld(pld: Int) = (pld==0)-> r_pld'.Qpld(pld+1) <>
		 (s_pld.Qpld(pld-1) + r_pld'.Qpld(pld+1));

Plant=Pu1||L0001||Pr1;
Supervisor=S0;

init
%%For asynchronous closed loop system.
hide({c_asc',c_desc',c_d0',c_d1',c_ext',c_pl1',c_pld',c_pu0',c_pu1',c_ret',
      c_up0',c_up1'},
allow({c_asc,c_desc,c_d0,c_d1,c_ext,c_pl1,c_pld,c_pu0,c_pu1,c_ret,c_up0,c_up1,
       c_asc',c_desc',c_d0',c_d1',c_ext',c_pl1',c_pld',c_pu0',c_pu1',c_ret',
       c_up0',c_up1'},
 comm({s_asc|r_asc->c_asc,s_desc|r_desc->c_desc,s_d0|r_d0->c_d0,s_d1|r_d1->c_d1,
       s_ext|r_ext->c_ext,s_pl1|r_pl1->c_pl1,s_pld|r_pld->c_pld,
       s_pu0|r_pu0->c_pu0,s_pu1|r_pu1->c_pu1,s_ret|r_ret->c_ret,
       s_up0|r_up0->c_up0,s_up1|r_up1->c_up1,
       s_asc'|r_asc'->c_asc',s_desc'|r_desc'->c_desc',s_d0'|r_d0'->c_d0',
       s_d1'|r_d1'->c_d1',s_ext'|r_ext'->c_ext',s_pl1'|r_pl1'->c_pl1',
       s_pld'|r_pld'->c_pld',s_pu0'|r_pu0'->c_pu0',s_pu1'|r_pu1'->c_pu1',
       s_ret'|r_ret'->c_ret',s_up0'|r_up0'->c_up0',s_up1'|r_up1'->c_up1'},
(Plant||Supervisor||Qasc(0)||Qdesc(0)||Qd0(0)||Qd1(0)||Qext(0)||Qpl1(0)||
 Qpld(0)||Qpu0(0)||Qpu1(0)||Qret(0)||Qup0(0)||Qup1(0))
)));

%% For synchrnous closed loop system.
%allow({c_asc,c_desc,c_d0,c_d1,c_ext,c_pl1,c_pld,c_pu0,c_pu1,c_ret,c_up0,c_up1},
% comm({s_asc|r_asc->c_asc,s_desc|r_desc->c_desc,s_d0|r_d0->c_d0,s_d1|r_d1->c_d1,
%      s_ext|r_ext->c_ext,s_pl1|r_pl1->c_pl1,s_pld|r_pld->c_pld,s_pu0|r_pu0->c_pu0,
%      s_pu1|r_pu1->c_pu1,s_ret|r_ret->c_ret,s_up0|r_up0->c_up0,s_up1|r_up1->c_up1},
%Plant||Supervisor
%));
\end{verbatim}
\section{mCRL2 model for the Pneumatic cylinder}\label{appendix-pc}
\begin{verbatim}
act
s_13,r_13,s_13',r_13',c_13,c_13',s_14,r_14,s_14',r_14',c_14,c_14',s_16,r_16,
s_16',r_16',c_16,c_16',s_23,r_23,s_23',r_23',c_23,c_23',s_24,r_24,s_24',
r_24',c_24,c_24',s_25,r_25,s_25',r_25',c_25,c_25',s_26,r_26,s_26',r_26',c_26,c_26';

proc
P0=r_13'.P1;
P1=(s_14'+s_16').P0;

Q0=r_23'.Q1;
Q1=r_25'.Q3+s_24'.Q2;
Q2=r_25'.Q3;
Q3=s_26'.Q0+r_23'.Q1;

Plant=P0||Q0;

%Note that for this case study 1 place queues are used.
%Because unbounded queues cause infinite states asynchronous
%closed loop system.

Q13(x13: Int) = (x13==0)-> r_13.Q13(x13+1) +
		 (x13==1)-> (s_13'.Q13(x13-1) + r_13.Q13(x13+1));

Q14(x14: Int) = (x14==0) -> r_14'.Q14(x14+1)+
		  (x14==1) ->( s_14.Q14(x14-1)+r_14'.Q14(x14+1));

Q16(x16: Int) = (x16==0) -> r_16'.Q16(x16+1) +
		  (x16==1) -> (s_16.Q16(x16-1)+r_16'.Q16(x16+1));

Q23(x23: Int) = (x23==0)-> r_23.Q23(x23+1) +
		 (x23==1)->( s_23'.Q23(x23-1)+r_23.Q23(x23+1));

Q25(x25: Int) = (x25==0)-> r_25.Q25(x25+1) +
		 (x25==1)->( s_25'.Q25(x25-1)+r_25.Q25(x25+1));

Q24(x24: Int) = (x24==0) -> r_24'.Q24(x24+1) +
		  (x24==1) -> (s_24.Q24(x24-1)+r_24'.Q24(x24+1));

Q26(x26: Int) = (x26==0) -> r_26'.Q26(x26+1) +
		  (x26==1) ->(s_26.Q26(x26-1)+r_26'.Q26(x26+1));

S0=s_13.S1;
S1=r_14.S0 + r_16.S2;
S2=s_23.S3;
S3=s_13.S4+r_24.S5;
S4=r_16.S3 +r_14.S6+ r_24.S7;
S5=s_13.S7;
S6=r_24.S8+s_25.S9;
S7=r_16.S5+r_14.S8;
S8=s_25.S9;
S9=s_13.S10+r_26.S0;
S10=r_16.S11 + r_26.S1 + r_14.S9;
S11=r_26.S2 + s_23.S3;

Supervisor=S0;
init
%For asynchronous closed loop system
hide({c_13',c_14',c_16',c_23',c_24',c_25',c_26'},
allow({c_13,c_13',c_14,c_14',c_16,c_16',c_23,c_23',c_24,c_24',c_25,
      c_25',c_26,c_26'},
    comm({s_13|r_13->c_13,s_13'|r_13'->c_13',s_14|r_14->c_14,s_14'|r_14'->c_14',
          s_16|r_16->c_16,s_16'|r_16'->c_16',s_23|r_23->c_23,s_23'|r_23'->c_23',
          s_24|r_24->c_24,s_24'|r_24'->c_24',s_25|r_25->c_25,s_25'|r_25'->c_25',
          s_26|r_26->c_26,s_26'|r_26'->c_26'},
        Plant||Q13(0)||Q14(0)||Q16(0)||Q23(0)||Q24(0)||Q25(0)||Q26(0)||Supervisor
    )))
  ;
%% For synchronous closed loop sytem
%allow({c_13,c_14,c_16,c_23,c_24,c_25,c_26},
%    comm({s_13|r_13->c_13,s_14|r_14->c_14,s_16|r_16->c_16,
%          s_23|r_23->c_23,s_24|r_24->c_24,s_25|r_25->c_25,
%          s_26|r_26->c_26},
%        Plant||Supervisor
%    ))%
%  ;
\end{verbatim}

\end{document}